\documentclass[fontsize=9pt, a4paper, twocolumn, DIV=20]{scrartcl}

\setkomafont{caption}{\footnotesize}     
\setkomafont{captionlabel}{\bfseries}    
\setkomafont{section}{\normalfont\large}
\setkomafont{subsection}{\normalfont\itshape}
\setkomafont{subsubsection}{\normalfont\itshape}
\setkomafont{paragraph}{\normalfont\itshape}
\addtokomafont{section}{\centering}

\usepackage{lipsum}

\usepackage{siunitx} 
\DeclareSIUnit{\rpm}{RPM}

\usepackage{upgreek} 
\usepackage[intlimits]{amsmath} 
\usepackage{subcaption} 
\usepackage{graphicx}   
\usepackage{flafter}    
\usepackage{booktabs}   
\usepackage[hidelinks]{hyperref}      
\usepackage[nameinlink,noabbrev,capitalise]{cleveref}     
\crefname{equation}{}{} 
\usepackage[maxcitenames=1, sorting=none, maxbibnames=99]{biblatex} 
\usepackage{amssymb}
\usepackage{amsthm}
\usepackage{makecell}
\usepackage{float}
\usepackage{multirow}

\usepackage{overpic}
\usepackage{tikz}
\usetikzlibrary{arrows.meta}

\usepackage[acronym]{glossaries}
\newacronym{uav}{UAV}{Unmanned Aerial Vehicle}
\newacronym{ekf}{EKF}{Extended Kalman Filter}
\newacronym{esc}{ESC}{Electronic Speed Controller}
\newacronym{ukf}{UKF}{Unscented Kalman Filter}
\newacronym{imu}{IMU}{Inertial Measurement System}
\newacronym{gps}{GPS}{Global Positioning System}
\newacronym{indi}{INDI}{Incremental Nonlinear Dynamic Inversion}
\newacronym{tcn}{TCN}{Temporal Convolutional Network}
\newacronym{rpm}{RPM}{Revolutions per Minute}
\newacronym{bem}{BEM}{Blade Element Momentum}
\newacronym{rmse}{RMSE}{Root Mean Square Error}
\newacronym{lasso}{LASSO}{Least Absolute Shrinkage and Selection Operator}
\newacronym{rps}{RPS}{revolutions per second}
\newacronym{ned}{NED}{North-East-Down}
\newacronym{ls}{LS}{Least Squares}
\newacronym{nrmse}{nRMSE}{normalized Root Mean Square Error}
\newacronym{rls}{RLS}{Recursive Least Squares}

\newcommand{\diff}[2]{\frac{\mathrm{d} #1}{\mathrm{d} #2}}

\glsdisablehyper

\title{Airspeed estimation for UAVs using only propeller feedback}
\author{Evangelos Ntouros\thanks{Correspondence: e.ntouros@tudelft.nl}\and
        Pavel Kelley\and
        Ewoud J. J. Smeur}

\defbibheading{bibliography}[\refname]{\section*{REFERENCES}}
\glsdisablehyper
\begin{document}

\date{}
\maketitle

\textbf{This work introduces a novel analytical model for estimating the airspeed of fixed-wing \glspl{uav} using solely propeller power and rotational speed measurements. The model can be used to replace Pitot-tube-based airspeed sensors, or contribute to redundancy in airspeed estimation. It does not require knowledge of the vehicle’s dynamic model and is computationally lightweight. It leverages power and rotational speed feedback, which is readily available from modern \glspl{esc}, thereby enabling seamless integration with existing systems and off-the-shelf components. A systematic approach is followed to derive the model structure based on least squares optimization and regularization techniques on \gls{bem} simulation, wind tunnel, and flight test datasets. The final model generalizes well achieving a \gls{nrmse} of 5\% on unseen flight data. The model parameters can be identified either offline, using flight logs with airspeed measurements, or in-flight, using a lightweight identification method based only on \gls{gps} velocity data. The resulting system provides a robust and computationally efficient solution for real-time airspeed estimation across diverse fixed-wing \gls{uav} platforms.}
\newline

\textbf{\textit{Keywords - UAV airspeed estimation, airspeed analytical redundancy, propeller feedback, in-flight parameter identification, BEM}}

\glsresetall

\section{INTRODUCTION}
\label{sec:introduction}

Airspeed information is critical for the operation of a \gls{uav}, as it is utilized in control algorithms for gain scheduling \cite{smeur_2020, poksawat_2018} and flight envelope protection.

Typically, the airspeed of a \gls{uav} is measured using a Pitot tube and a set of differential pressure sensors. This configuration measures the dynamic pressure of the airflow, which is the difference between the total pressure captured by a forward-facing tube and the static pressure of the surrounding environment. However, Pitot tubes are particularly vulnerable to environmental factors such as icing or clogging, caused by dust and water, while the pressure sensors are susceptible to temperature fluctuations. 

One way to increase the reliability of the airspeed measurement is by adding different sensors, thus introducing physical redundancy. For example, Makaveev et al. \cite{makaveev_2023} proposed a microphone-based setup that estimates airspeed by analyzing the power spectra of wall-pressure fluctuations generated by the turbulent boundary layer over the vehicle’s surface during flight. A feed-forward single-layer neural network is then employed to predict the airspeed based on the power spectra of the microphone signals. Li et al. \cite{li_2019} proposed the integration of a mass-flow sensor that feeds an indoor localization algorithm with robust velocity updates while \cite{thielicke_2021} utilized an ultrasonic anemometer on a \gls{uav} for accurate wind measurements. While physical redundancy can mitigate the risk of airspeed sensor failure, it increases the weight and cost of the \gls{uav}. 

An alternative to adding additional physical sensors is analytical redundancy, which offers a cost-effective solution and enables operation even in the absence of an airspeed sensor. Some methods exploit the dynamic model of the \gls{uav} to derive the estimate. For example, Fravolini et al. \cite{fravolini_2012} demonstrated that, when the state of the system is known, except for the airspeed, the angle of attack results quadratic in airspeed, thereby enabling to solve for it. Ganguli et al. \cite{Ganguli_2013} approximated airspeed using a model of the aircraft and the current signal from the servo motors actuating on the elevons of the \gls{uav}. The authors in \cite{guo_2025} employed aerodynamic and thrust modeling techniques to estimate airspeed, and subsequently incorporating it into the control structure of a solar-powered \gls{uav}.

Data-driven approaches have also been explored. For instance, researchers in \cite{book_samy_2011, Gururajan_2013} applied Machine Learning techniques to estimate airspeed for the purpose of hardware failure detection. Lim et al. \cite{Lim_2022} utilized a \gls{tcn}, using \gls{imu} data, elevator input, and airflow angle measurements, to generate synthetic airspeed measurements, which were subsequently fused using a \gls{ukf}.

Other methods rely on sensor fusion. Rhudy et al. \cite{Rhudy_2015} fused measurements from onboard sensors within a nonlinear Kalman filter, simultaneously estimating both the airspeed and the local wind vector without requiring a dynamic model. However, their approach required the extra use of vanes to measure the relative wind, which are typically not part of most \gls{uav} setups. Similarly, Guo et al. \cite{guo_2018} developed an \gls{ekf}-based method that integrates data from an \gls{imu}, \gls{gps}, and wind vanes, while further formally proving the observability of the nonlinear system describing the airspeed kinematics.

We propose a novel airspeed estimation approach purely based on propeller power and rotational rate measurements, without the need of a vehicle dynamics model or any complex computations. The method leverages the physical relationship between the freestream velocity, propeller rotational speed, and propeller power. Modern \glspl{esc} natively provide feedback on input power to the ESC-motor-propeller system and on the rotational speed of the propeller, so the proposed approach is compatible with off-the-shelf components and can be seamlessly integrated into existing platforms without hardware modifications. Additionally, we present an in-flight model parameter identification procedure that requires only earth-frame velocity data, which are typically available on most \gls{uav} platforms via \gls{gps} measurements. This removes the dependency on existing pre-flight airspeed data, thereby enabling autonomous, self-configuring deployment. 

\section{METHODOLOGY}
\label{sec:theory}

To estimate the airspeed of a \gls{uav} from propeller power and rotational speed, a relationship between these variables must be established. In the following subsections, we address this problem by proposing two alternative model categories and present the optimization method utilized to extract their corresponding terms from data.

In our approach, we assume nominal flight conditions characterized by small angle of attack $\alpha$ and sideslip angle $\beta$ such that the propeller operates under approximately axial flow conditions. Nevertheless, we demonstrate that the airspeed estimate maintains reasonable accuracy even during more aggressive flight, where violations of these assumptions occur.

\subsection{Indirect airspeed model}
\label{subsec:indirect}

Using the Buckingham-$\pi$ theorem \cite{anderson2016}, propeller power is defined with the help of the non-dimensional power coefficient $C_P$ as
\begin{equation}
    P = C_P(J, M_{\text{tip}}, Re_{\text{tip}})\, \rho_a n^3 D^5, 
    \label{eq:P}
\end{equation}
where $J=\frac{V_a}{nD}$ is the advance ratio, $V_a$ the freestream velocity, $D$ the propeller diameter, $n$ the rotational speed in \gls{rps}, $\rho_a$ the air density, $M_{tip}$ the tip Mach number, and $Re_{tip}$ the tip Reynolds number. In the case of a \gls{uav} flight we can assume that compressibility and viscous effects are negligible, thus we may approximate $C_P(J,M_{tip}, Re_{tip}) = C_P(J)$.

We define $\mathcal{J} \subset \mathbb{R}$ as the set of all advance ratios $J$ within the flight envelope. Then we define the power coefficient as the mapping from $\mathcal{J}$ to $\mathcal{C_P}$, seen in \cref{eq:Cp-vs-J}, where $\mathcal{C_P} \subset \mathbb{R}$. To calculate $V_a$ we need to model the inverse mapping, \cref{eq:J-vs-Cp}, and then apply $V_a = nDJ$. Hence, we call this an \emph{indirect} approach.
\begin{subequations}
\begin{align}
    C_P &: \mathcal{J} \to \mathcal{C_P} \label{eq:Cp-vs-J} \\
    C_P^{-1} = J &: \mathcal{C_P} \to \mathcal{J} \label{eq:J-vs-Cp}
\end{align}
\end{subequations}

Modeling the mapping in \cref{eq:J-vs-Cp} using an analytical expression, requires \cref{eq:Cp-vs-J} to be invertible, thus bijective. Existing literature \cite{gur_2024,redgrave2015,brandt_2011}, and datasets from the current work presented in \cref{fig:Cp-J_fit_all} suggest that \cref{eq:Cp-vs-J} is a concave function, hence not bijective.
Consequently, deriving a globally valid analytical expression for the airspeed across all advance ratios is not feasible.
The solution is to divide $C_P(J)$ into invertible sections and treat them separately.
Since there is no apriori known formula of the function we define below the generic inversion criterion for $C_P(J)$ or equivalently the data \emph{selection criterion} for a feasible model fit on $J$--$C_P$ data points.

\paragraph{Selection criterion and critical operating points}
Let $\mathcal{J}_I \subset \mathcal{J}$ be a connected interval over which the derivative $\diff{C_P}{J}$ does not change sign.
Then $C_P$ is said to be monotonic on $\mathcal{J}_I$, and thus invertible. In this case, it is possible to construct an analytical expression for $J$, and consequently for $V_a$, over this interval. For the remainder of this work, we refer to this as the \emph{selection criterion}, and to the advance ratio values, which define the endpoints of the interval $\mathcal{J}_I$, as the \emph{critical operating points}.

\subsection{Direct airspeed model}
\label{subsec:direct}

Attempting to calculate $V_a$ by estimating $J$ from $C_P$ measurements aligns with the literature but we impose a certain structure to the model, potentially constraining the optimization algorithm, described subsequently in \cref{subsec:optimization}, in extracting the model features from a smaller subset of possible representations.

A second, more \emph{direct} approach seeks to mitigate this issue by estimating airspeed directly from propeller power and rotational speed measurements. In this case we define $\mathcal{V}$ and $\Omega$ as the sets of all airspeeds $V_a$ and propeller rotational speeds $\omega$, respectively, within the flight envelope; the power required by the propeller under these operating conditions belongs in the set $\mathcal{P} \subset \mathbb{R}$. The airspeed can then be expressed in terms of $P$ and $\omega$:
\begin{equation}
    V_a \colon \mathcal{P} \times \Omega \to \mathcal{V} \label{eq:Va-vs-Pw}.
\end{equation}
The selection criterion outlined in \cref{subsec:indirect} also applies to this direct modeling approach.

\subsection{Optimization method}
\label{subsec:optimization}

In \cref{subsec:indirect} and \cref{subsec:direct}, we defined the airspeed using a generic representation. To extract analytical formulas from input–output datasets, we employ the \gls{lasso} method within a $k$-fold cross-validation framework \cite{brunton-kutz2022}. At its core, the algorithm solves a least squares problem while penalizing model complexity. It promotes model sparsity by selecting only the dominant terms, thereby avoiding overfitting and improving generalizability. A tunable parameter $\lambda$ controls the degree of sparsity, with larger values yielding sparser models.
 
We construct the input matrix to include a wide range of candidate features formed by various combinations of the input variables. The \gls{lasso} method then fits this over-parameterized model to the data for a range of $\lambda$ values, evaluating its performance through a cross-validated error metric. Finally, the value of~$\lambda$, and thus the corresponding model terms, is selected accordingly such that a low cross-validated error is achieved while maintaining a model that is as simple as possible.

\section{DATA COLLECTION AND EXPERIMENTAL SETUPS}
\label{sec:datasets}

\subsection{BEM dataset}
\label{subsec:bem_dataset}
We follow the BEM implementation proposed by Ning \cite{ning_2014} to construct a $(P,\omega, V_a)$ dataset.
The simulation input is the propeller parameters, the freestream velocity, which is varied from \qty{0}{} to \qty[per-mode=symbol]{30}{\meter\per\second}, and the propeller rotational speed, which is varied from \qty{1000}{} to \qty{10000}{\rpm}. The simulation output is the propeller power.
The propeller parameters are measured from the propeller of the test vehicle, presented in \cref{subsec:flight_dataset}, thereby enabling direct comparison of the results.
For these measurements, we employed an approach similar to that described in \cite{mcCrink2017}.
\Cref{fig:Cp-J_fit_all} illustrates the $C_P$--$J$ data points of this dataset.

\subsection{Flight test datasets}
\label{subsec:flight_dataset}
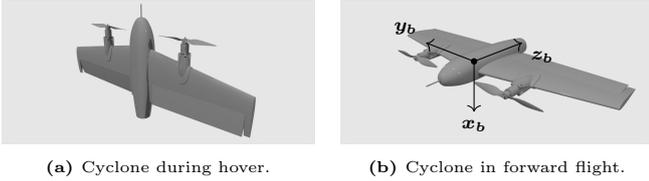
\begin{figure}[h]
    \centering

    \begin{subfigure}[t]{0.48\columnwidth}
        \centering
        \includegraphics[width=\columnwidth]{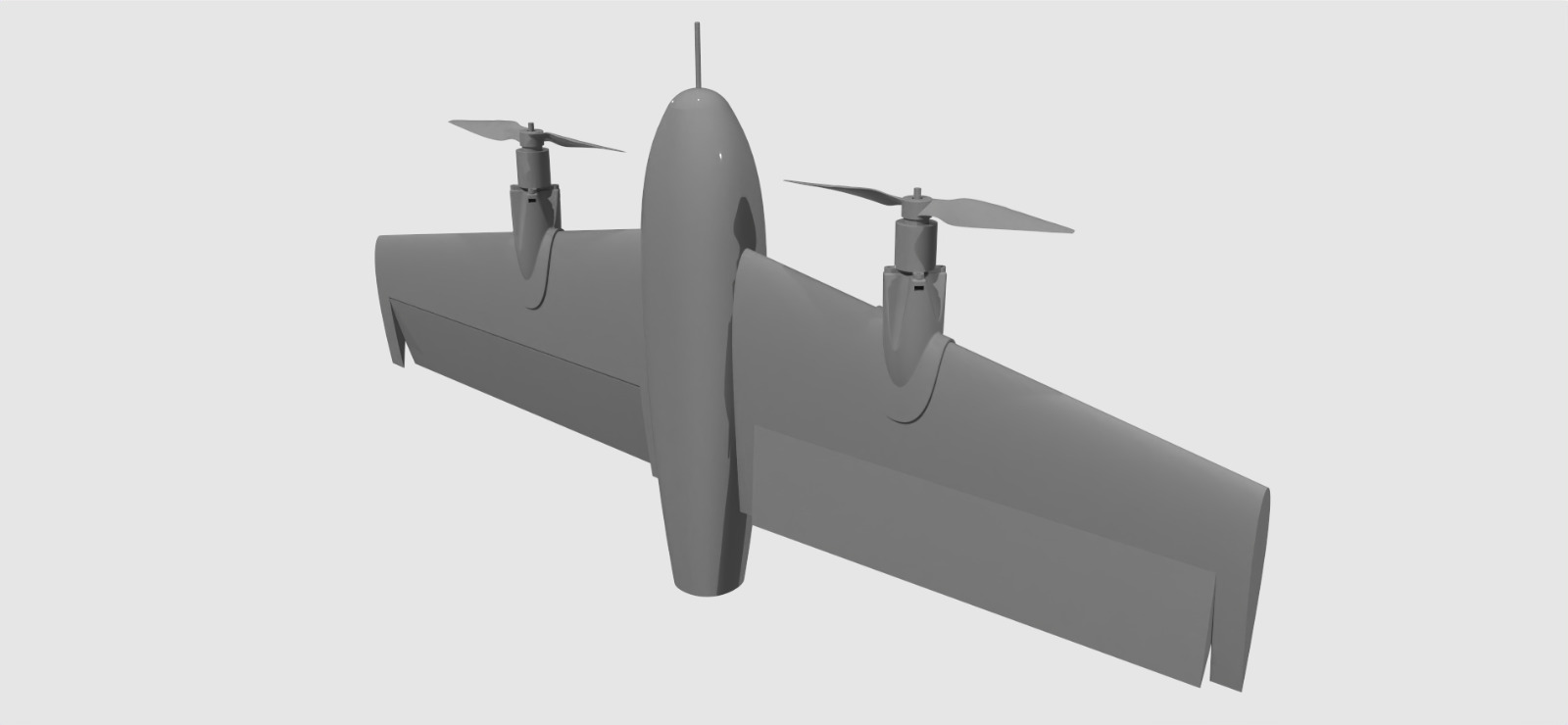}
        \caption{Cyclone during hover.}
        \label{photo_cyclone_hover}
    \end{subfigure}
    \hfill
    \begin{subfigure}[t]{0.48\columnwidth}
        \centering
        \begin{overpic}[width=\columnwidth]{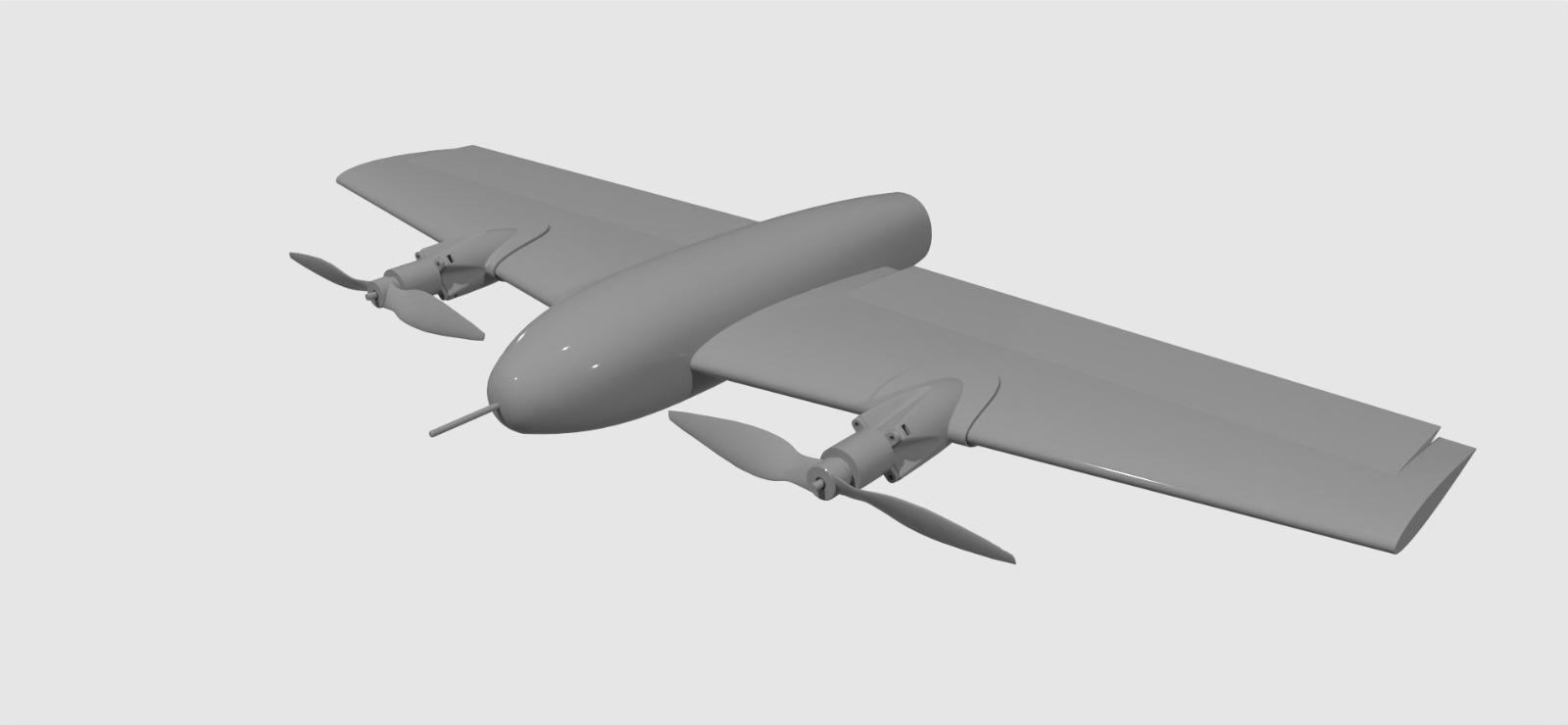}
            \begin{tikzpicture}[overlay, x=1pt, y=1pt]
                \coordinate (O) at (50, 31);
                \draw[fill=black] (O) circle (1.0pt);
                \draw[->, black] (O) -- ++(-18, 7) node[above left] {\footnotesize $\boldsymbol{y_b}$};
                \draw[->, black] (O) -- ++(0, -19) node[below] {\footnotesize $\boldsymbol{x_b}$};
                \draw[->, black] (O) -- ++(18, 7) node[below right] {\footnotesize $\boldsymbol{z_b}$};
            \end{tikzpicture}
        \end{overpic}
        \caption{Cyclone in forward flight.}
        \label{photo_cyclone_fw}
    \end{subfigure}

    \caption{The Cyclone tailsitter \gls{uav}.}
    \label{photo_cyclone}
\end{figure}

To obtain real flight data, we conducted flight tests using Cyclone, the tailsitter \gls{uav} depicted in \cref{photo_cyclone_hover} and \cref{photo_cyclone_fw} in hover and forward flight attitude, respectively.
It is equipped with a Pitot tube to obtain airspeed measurements and a \gls{gps} module that provides earth-frame velocity data.

We conduct manual flights with turns, speed, and altitude variations so that they better reflect generic operational scenarios.
We apply a correction to the airspeed measured by the Pitot tube $V_{a,\textrm{pitot}}$ to correctly express the freestream velocity at the propeller, which is offset from the vehicle’s longitudinal axis by a distance $l = 0.24\,\unit{\meter}$.
We collect feedback from the right \gls{esc}, so the appropriate correction is
\begin{equation}
    V_a = V_{a,\textrm{pitot}} - \Omega_xl,
    \label{eq:airspeed_correction}
\end{equation}
where $\Omega_x$ denotes the angular velocity of the vehicle about its body-$x$ axis, as illustrated in \cref{photo_cyclone_fw}.

\subsubsection{Electro-mechanical efficiency and propeller power}

The \gls{esc} provides feedback for input voltage $V$, current $I$, and rotational speed $\omega$.
The input power to the ESC-motor-propeller system is then calculated by $P_\textrm{in} = VI$.
Based on that, propeller power $P$ is computed with
\begin{align}
    P = \eta P_\textrm{in},
    \label{eq:prop_power}
\end{align}
where the electro-mechanical efficiency of the ESC-motor system $\eta$, accounts for power losses, namely \gls{esc} switching losses, resistive losses in the motor windings, iron core losses in the motor, as well as frictional losses.
It does not include the aerodynamic efficiency of the propeller.
We assume that the efficiency $\eta$ is constant with respect to current, temperature, and rotational speed changes.

\begin{figure}[h]
    \centering
    \includegraphics[width=0.86\columnwidth]{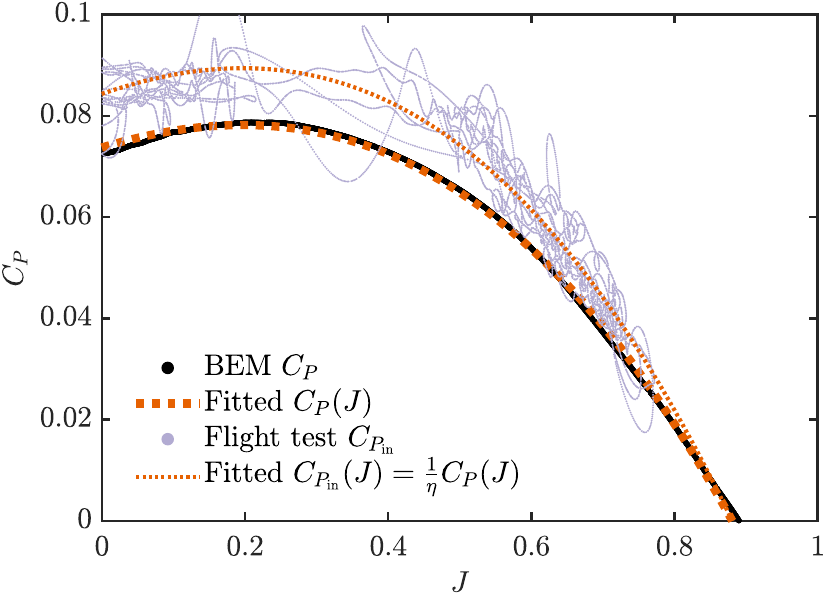}
    \caption{Propeller power coefficient from \gls{bem} data and \gls{esc} input power coefficient from flight data.
The electro-mechanical efficiency of the ESC-motor system relates these two coefficients.}
    \label{fig:Cp-J_eta}
\end{figure}

We estimate the electro-mechanical efficiency of the ESC-motor system, by analysing the \gls{bem} and the flight test data, depicted in \Cref{fig:Cp-J_eta}.
The \gls{bem} data refer to the propeller power coefficient $C_P = \frac{P}{\rho_a n^3 D^5}$. For the flight test data, propeller power is not directly available so we define the \gls{esc} input power coefficient as $C_{P_\textrm{in}} = \frac{P_\mathrm{in}}{\rho_a n^3 D^5} = \frac{1}{\eta}C_P$, which is also depicted in the same figure.

We estimate the efficiency $\eta$ as follows. First, we fit a third-order polynomial to the BEM-derived $C_P$--$J$ data, yielding the curve $C_P(J) = 0.074 + 0.043 J - 0.092 J^2 - 0.059 J^3$, as depicted in the figure. Next, we scale this polynomial by a parameter $1/\eta$, getting the expression $\frac{1}{\eta} C_P(J)$, which we fit to the flight test-derived $C_{P_\textrm{in}}$--$J$ data resulting in the fit shown in the same figure. Accordingly, we obtain $1/\eta = 1.15$, and thus, $\eta = 0.87$. 

Propeller power for the flight test is then calculated by \cref{eq:prop_power} and the $(P,\omega, V_a)$ dataset is obtained.
We collect 2 flight test datasets, the \emph{training} one for fitting the model and the \emph{test} one for testing its performance on unseen data.
\Cref{fig:Cp-J_fit_all} shows the $C_P$--$J$ data points of the training dataset.

\subsection{Wind tunnel dataset}
\label{subsec:wt_datasets}

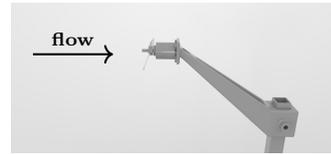
\begin{figure}[h]
    \centering
    \begin{overpic}[width=0.5\columnwidth]{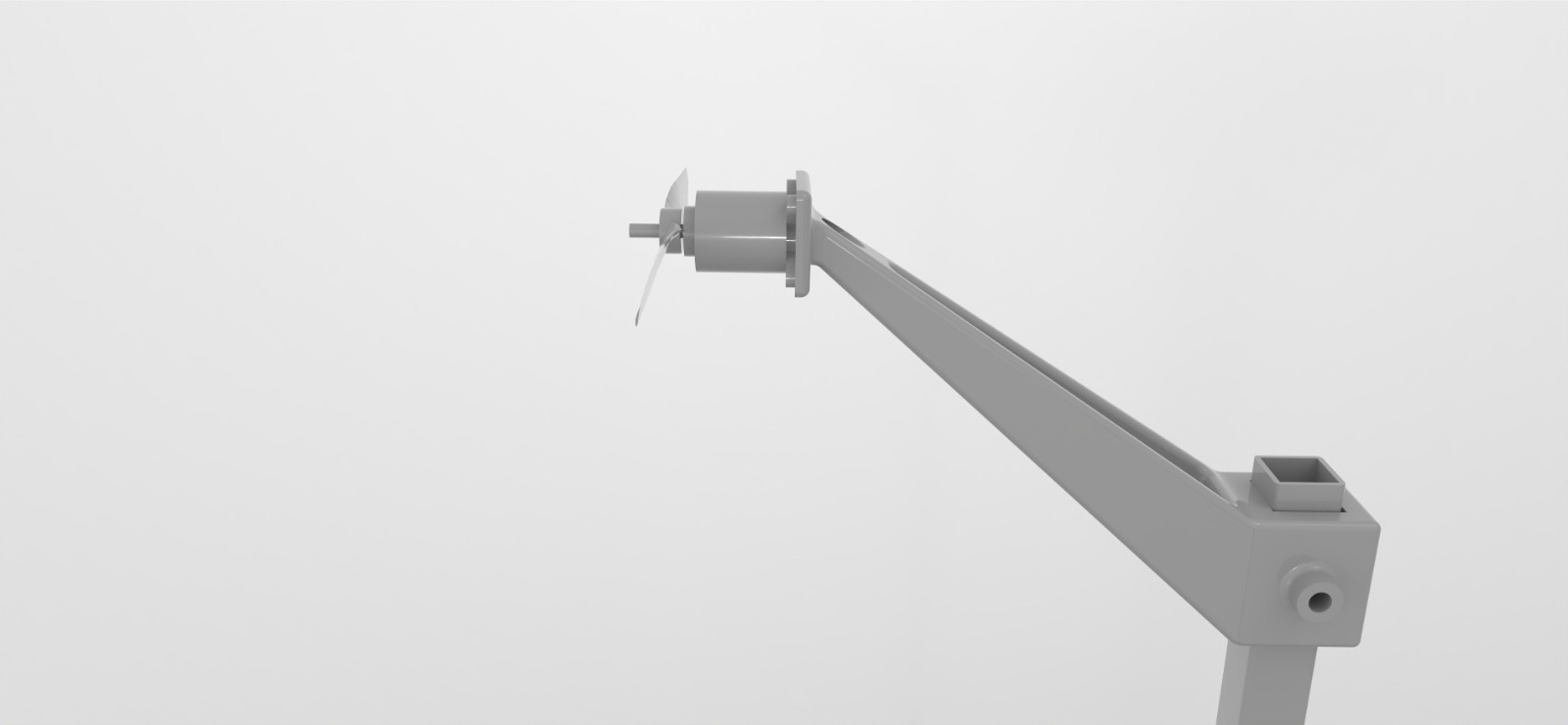}
        \begin{tikzpicture}[overlay, x=1pt, y=1pt]
            \draw[->, thick, black] (8,37) -- ++(30,0) node[midway, above] {\footnotesize \textbf{flow}};
        \end{tikzpicture}
    \end{overpic}
    \caption{Wind tunnel experimental setup.}
    \label{photo_wt_setup}
\end{figure}

\begin{figure}[h]
    \centering
    \includegraphics[width=0.86\columnwidth]{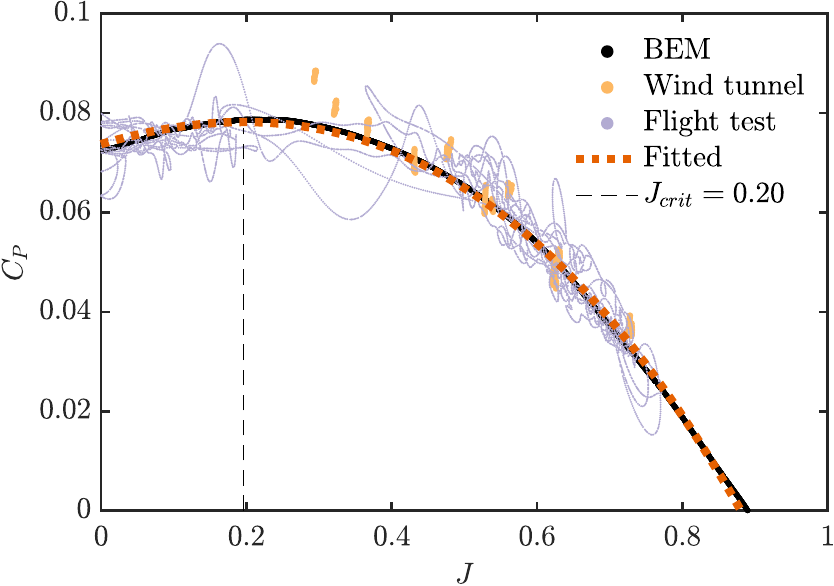}
    \caption{Propeller power coefficient as a function of advance ratio for the three training datasets. The fitted curve on \gls{bem} data is used to evaluate the selection criterion.}
    \label{fig:Cp-J_fit_all}
\end{figure}

We also collect data from wind tunnel experiments. \Cref{photo_wt_setup} shows the schematic of the experimental setup, which includes the same ESC-motor-propeller system as the Cyclone \gls{uav}. The setup ensures that the wind tunnel’s flow is axial to the propeller. The motor is driven with constant throttle input commands ranging from 10\% to 80\%, with an upper rotational speed limit of \qty{10000}{rpm} to ensure thermal protection. \gls{esc} feedback is recorded at three freestream velocities: $V_a =$ \qty[per-mode=symbol]{10}{\meter\per\second}, \qty[per-mode=symbol]{15}{\meter\per\second}, and \qty[per-mode=symbol]{18}{\meter\per\second}. Using \cref{eq:prop_power}, we estimate the power consumed by the propeller and construct the $(P, \omega, V_a)$ dataset. \Cref{fig:Cp-J_fit_all} shows the $Cp$--$J$ data points from this dataset. 

\section{DERIVING THE AIRSPEED MODEL}
\label{sec:model}

In this section, we derive an explicit airspeed formula for both the direct and the indirect model categories. The optimization method discussed in \cref{subsec:optimization} is used with the datasets described in \cref{sec:datasets} to derive the analytical model structures.

\subsection{BEM}
\label{subsec:bem}

\begin{figure}[h]
    \centering
    \includegraphics[width=0.86\columnwidth]{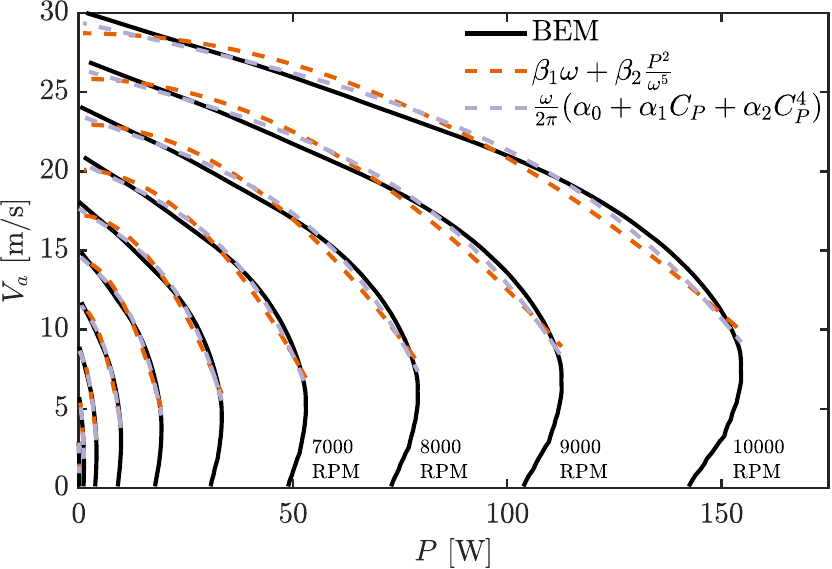}
    \caption{Direct and indirect model fits for $J>J_\text{crit}$ on the \gls{bem} dataset.}
    \label{fig:bem_Va-P-wconst_fit}
\end{figure}

We begin by applying the selection criterion to the BEM-generated $C_p$--$J$ data points. We use the cubic polynomial fit $C_P(J) = 0.074 + 0.043 J - 0.092 J^2 - 0.059 J^3$, previously obtained in \cref{subsec:flight_dataset}, to find the critical operating points as the roots of $\diff{C_P}{J}$. Two solutions are obtained; one is negative and is thus discarded. The other is the critical operating point $J_{\text{crit}} = 0.20$. This value defines two distinct operational regimes in which airspeed estimation is separately feasible. Advance ratio values in the interval $[0, J_{\text{crit}}]$ correspond primarily to hovering conditions, characterized by large angle of attack, while for $J > J_{\text{crit}}$ they correspond to nominal forward flight conditions. This study focuses on the latter; therefore, our analysis is restricted to $J > J_{\text{crit}}$, as illustrated in \cref{fig:bem_Va-P-wconst_fit}.

Applying the optimization method, we extract the dominant features for the two model categories, resulting in the following representations:
\begin{subequations}
\begin{align}
\text{\textit{Indirect:}} \quad & V_a(C_P, \omega) = \frac{\omega}{2\pi}D(\alpha_0 + \alpha_1 C_P + \alpha_2 C_P^4)
\label{eq:Va-vs-Cp_steady} \\
\text{\textit{Direct:}} \quad & V_a(P, \omega) = \beta_1 \omega + \beta_2 \frac{P^2}{\omega^5}
\label{eq:Va-vs-Pw_steady}
\end{align}
\label{eq:Va_bem}%
\end{subequations}%
We observe the direct model $V_a(P,\omega)$ getting a slightly simpler form than the indirect one, $V_a(C_P,\omega)$. \Cref{fig:bem_Va-P-wconst_fit} illustrates the corresponding model fits. Overall both models demonstrate highly accurate fits, with small deviations near $J_\mathrm{crit}$. This is expected, as the curves become very steep in that region, making a parsimonious model more difficult to fit. In \Cref{tab:pred} we present the fitting accuracy metrics \gls{rmse} and \gls{nrmse}. To get the \gls{nrmse} value we normalize by the range of the airspeed values, \qty[per-mode=symbol]{30}{\meter\per\second}. This enables consistent comparisons across the different datasets. \Cref{tab:coeffs} presents the coefficient values for the two identified models.

This analysis reveals the structure of the two model categories; however, the results should be interpreted with caution, as the models were derived from simulated data that do not include all the complexities of the real-world system. Further analysis follows to investigate the validity of the models on real-world scenarios.

\subsection{Wind tunnel}
\label{subsec:wt}

\begin{figure*}[h]
    \centering
    \begin{subfigure}[t]{0.86\columnwidth}
        \includegraphics[width=\columnwidth]{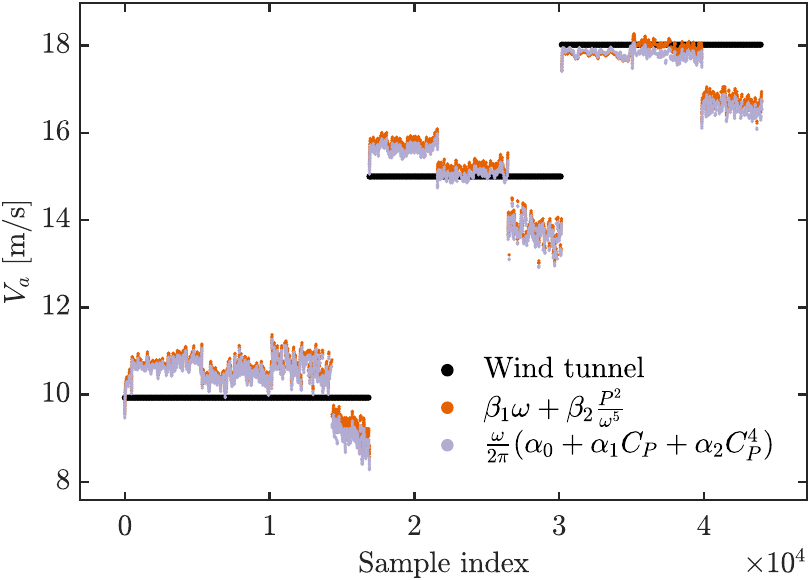}
        \caption{Direct and indirect model fits for $J>J_\mathrm{crit}$.}
        \label{fig:wt_Va_fit}
    \end{subfigure}
    \begin{subfigure}[t]{0.86\columnwidth}
        \includegraphics[width=\columnwidth]{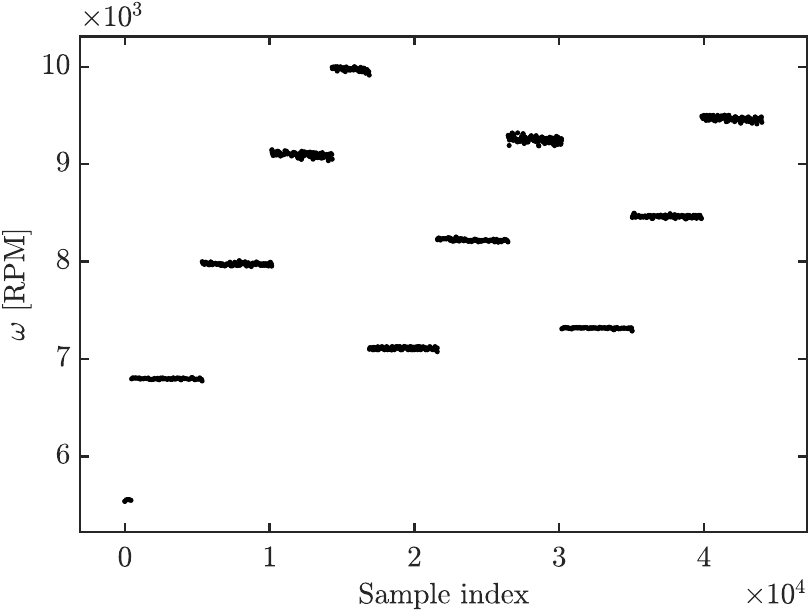}
        \caption{Propeller rotational speed.}
        \label{fig:wt_rpm}
    \end{subfigure}
    \caption{Wind tunnel dataset.}
    \label{fig:wt}
\end{figure*}

To validate the proposed model structures using real data, we continue the analysis on the wind tunnel dataset. We apply the selection criterion and fit the models to the wind tunnel data. \Cref{fig:wt} presents a comparison of the two fits. Data sections excluded by the selection criterion have been omitted from the plot. \Cref{tab:pred} presents the fitting error metrics. For the \gls{nrmse} metric we normalize by the range of the freestream velocity, \qty[per-mode=symbol]{8}{\meter\per\second}. We observe an accurate fit in the mid-range RPM region, with a slight degradation at higher RPM values. Moreover the accuracy at \qty[per-mode=symbol]{10}{\meter\per\second} is more consistent. The results agree with the findings and trends of the \gls{bem} analysis, further supporting the validity of the previously established model structures. The corresponding coefficients can be found in \cref{tab:coeffs}.

\subsection{Flight test}
\label{subsec:flight}

Next, we validate the proposed model structures using the flight test dataset. After applying the selection criterion, the models are fit to the flight data. Both models demonstrate a high degree of fitting accuracy, as shown in \cref{tab:pred}. The \gls{nrmse} values are obtained by normalizing with the airspeed values range, \qty[per-mode=symbol]{18.3}{\meter\per\second}. These results extend the validity of the derived model structures to real flight conditions, showing that they effectively capture the core dynamics of the propeller operation. The corresponding model coefficients are listed in \cref{tab:coeffs}.

\section{IN-FLIGHT MODEL COEFFICIENTS IDENTIFICATION}
\label{sec:inflight}
 
In \cref{sec:model}, we identified two candidate airspeed models, trained on datasets with airspeed measurements. Here, we present a method for identifying the model coefficients without relying on such measurements, but instead using earth-frame velocity data.

To facilitate this, we leverage the derived model structures \cref{eq:Va_bem}. We extend the training dataset from \cref{subsec:flight_dataset} with earth-frame velocity data from \gls{gps} measurements in the \gls{ned} frame,  
$
\boldsymbol{V} = 
\begin{bmatrix}
V_{N} & V_{E} & V_D
\end{bmatrix}^T
$.
Additionally we assume a constant wind field, both spatially and in time in the horizontal plane, expressed in \gls{ned} as
$
\boldsymbol{V}_w = 
\begin{bmatrix}
V_{w_N} &
V_{w_E} &
0
\end{bmatrix}^T.
$
Then, under the assumption that $\beta \approx 0$, we formulate the least squares optimization problem as
\begin{equation}
    \boldsymbol{V_g} = \begin{bmatrix}
        V_{N} \\
        V_{E}
    \end{bmatrix} \\
    = \begin{bmatrix}
        \hat{V_a} \cos(\gamma) \cos(\psi) + V_{w_N} \\
        \hat{V_a} \cos(\gamma) \sin(\psi) + V_{w_E}
    \end{bmatrix} \\,
\label{eq:gs_lr}
\end{equation}
where $\hat{V_a}$ denotes the direct or the indirect model, $V_a(P,\omega)$ or $V_a(C_P, \omega)$, respectively. Assuming a small roll angle, the flight path angle, $\gamma$, is calculated by
\begin{equation}
    \gamma = \arcsin\left(\frac{V_D}{\| \boldsymbol{V} \|}\right).
    \label{eq:gamma}
\end{equation}
The complete formulation for the direct case is presented in \cref{eq:gs_lr_direct}, whereas for the indirect in \cref{eq:gs_lr_indirect}.
\begin{subequations}
\begin{align}
    \begin{bmatrix}
    \begin{bmatrix} \omega & \frac{P^2}{\omega^5}\end{bmatrix}\cos(\gamma)\cos(\psi) & 1 & 0 \\
    \begin{bmatrix} \omega & \frac{P^2}{\omega^5}\end{bmatrix}\cos(\gamma)\sin(\psi) & 0 & 1
    \end{bmatrix}
    \begin{bmatrix}
    \beta_1 \\
    \beta_2 \\
    V_{w_N} \\
    V_{w_E}
    \end{bmatrix}
    &=
    \begin{bmatrix}
    V_{N} \\
    V_{E}
    \end{bmatrix}
    \label{eq:gs_lr_direct} \\
    \begin{bmatrix}
    \begin{bmatrix} 1 & C_P & C_P^4 \end{bmatrix}\cos(\gamma)\cos(\psi) & 1 & 0 \\
    \begin{bmatrix} 1 & C_P & C_P^4 \end{bmatrix}\cos(\gamma)\sin(\psi) & 0 & 1
    \end{bmatrix}
    \begin{bmatrix}
    \alpha_0 \\
    \alpha_1 \\
    \alpha_2 \\
    V_{w_N} \\
    V_{w_E}
    \end{bmatrix}
    &=
    \begin{bmatrix}
    V_{N} \\
    V_{E}
    \end{bmatrix}
    \label{eq:gs_lr_indirect}
\end{align}
\end{subequations}

Consistently, the proposed method yields an accurate fit for the direct model. In contrast, the indirect model exhibits inferior performance, as shown in \cref{tab:pred}. The error metric is normalized with the airspeed range, \qty[per-mode=symbol]{24.7}{\meter\per\second}. \Cref{tab:coeffs} shows the identified coefficients of the two models. It is evident that for the direct model, the coefficients remain consistent across the different datasets and fitting methods. However, for the indirect model, the coefficients vary significantly in the current case. This further highlights the issue of the reduced fitting accuracy.

Fitting the model using airspeed versus using earth-frame velocity data differs only in the additional assumptions required in the latter case. Thus, the discrepancy observed in the fitting performance of the indirect model between these two approaches can be attributed to reduced robustness under violations of the underlying assumptions, which are present in the flight dataset.

We note that if the electro-mechanical efficiency $\eta$ is not known, the input power at the \gls{esc} can be used instead of the propeller power in the airspeed model. Then the coefficient identification procedure will estimate parameters that implicitly incorporate the unknown efficiency $\eta$.

\section{MODEL EVALUATION ON UNSEEN DATA}
\label{sec:prediction}

In this section, we assess the extrapolation capability of the two models by evaluating their predictive accuracy on unseen data. To this end, the test dataset described in \cref{subsec:flight_dataset} is employed, comprising a vertical take-off, a transition to forward flight, and, at the end, a final transition back to hover for landing. 

Since the airspeed is not known \emph{a priori}, the selection criterion cannot be directly evaluated for real-time deployment of the method. Rigorous selection is not feasible in this context; therefore, we suggest using the angle of attack estimation 
\begin{equation}
    \alpha = \theta - \gamma = \theta - \arcsin(\frac{V_D}{\| \boldsymbol{V} \|})
    \label{eq:alpha}
\end{equation}
as an alternative criterion, where $\theta$ is the \gls{uav} pitch angle.

\begin{figure*}[h]
    \centering
    \begin{subfigure}[t]{0.86\columnwidth}
        \includegraphics[width=\columnwidth]{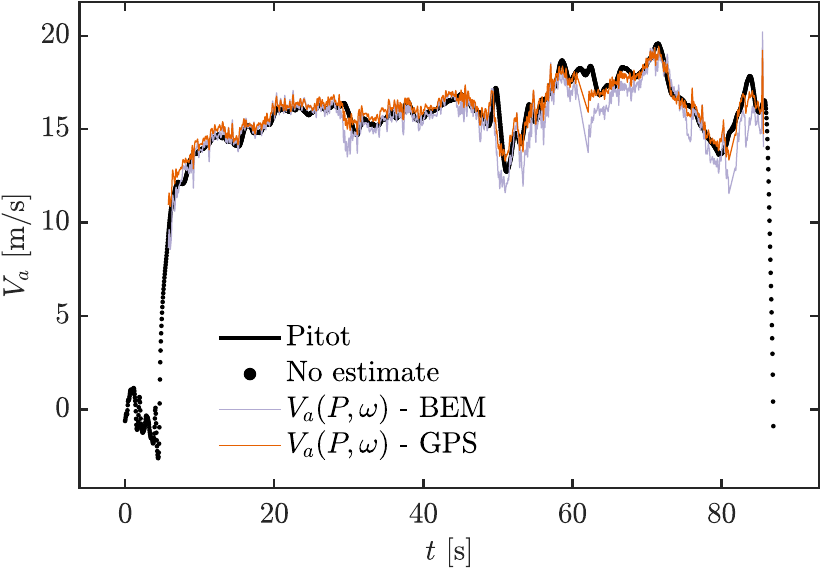}
        \caption{Airspeed prediction. Comparison of $V_a(P, \omega) = \beta_1 \omega + \beta_2 \frac{P^2}{\omega^5}$ identified by \gls{bem} data versus by flight data with GPS earth-frame velocity measurements.}
        \label{fig:flight_Va_predict}
    \end{subfigure}
    \begin{subfigure}[t]{0.955\columnwidth}
        \includegraphics[width=\columnwidth]{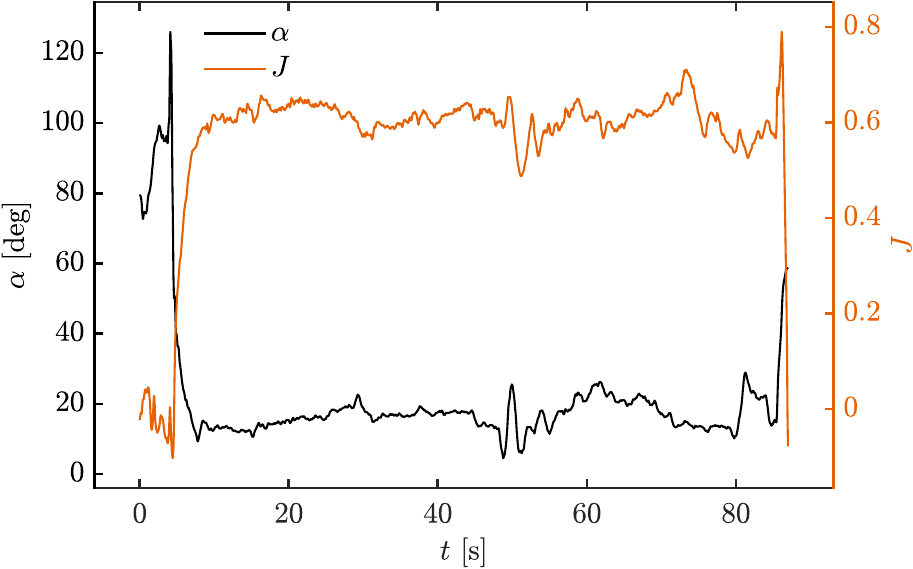}
        \caption{Angle of attack and advance ratio.}
        \label{fig:flight_alpha_J}
    \end{subfigure}
    \caption{Direct model evaluation on the test dataset.}
\end{figure*}

\Cref{fig:flight_alpha_J} presents the advance ratio alongside the estimate of the angle of attack during the test flight. It is evident that the principal function of the selection criterion is to effectively isolate the forward flight segment of the tailsitter flight while excluding the hover phase. The distinction between hover and forward flight phases for the tailsitter can be intuitively characterized by the angle of attack. Values near 90\si{\degree} indicate hover, whereas smaller values correspond to forward flight. The transition between these two phases can be realised as a distinct shift from higher to lower angle of attack values, and \emph{vice versa}, defined by a threshold value for the angle of attack, $\alpha_\textrm{th}$.

A function of the selection criterion, not captured by the angle of attack alternative criterion, is to exclude the operating conditions where the \gls{uav} operates at low speed and then rapidly increases throttle. The propeller typically accelerates faster than the vehicle and momentarily the vehicle may enter the $J<J_\textrm{crit}$ regime. This issue can be mitigated by using the airspeed estimates only above an empirically defined low airspeed threshold.

We select $\alpha_\textrm{th} = 25\unit{\degree}$ and test both model structures against the unseen data, utilizing the various sets of coefficients from \cref{tab:coeffs}. The corresponding detailed results are presented in \Cref{tab:pred}. When the model coefficients are identified using airspeed data, both the direct and indirect approaches demonstrate comparable predictive accuracy, as seen in \cref{tab:pred}. The airspeed range used for normalizing the prediction error is \qty[per-mode=symbol]{10}{\meter\per\second}. Small discrepancies are observed, particularly in the second half of the flight, where large angle of attack occurs. This is consistent with our methodology, which assumes small angle of attack. Nevertheless, the model demonstrates a notable level of robustness as also depicted in \Cref{fig:flight_Va_predict}. On the other hand, when earth-frame velocity data are used in the in-flight identification procedure, the direct model exhibits superior predictive performance. This is expected, as the indirect model had already demonstrated inferior fitting performance in \cref{sec:inflight}.

Following the preceding analysis, we adopt the model $V_a(P, \omega) = \beta_1 \omega + \beta_2 \frac{P^2}{\omega^5}$ as the proposed method for airspeed estimation. \Cref{fig:flight_Va_predict} compares the prediction of the BEM-trained model with that of the in-flight earth-frame velocity-trained model. Notably, satisfactory performance is also achieved using the model trained solely on \gls{bem} simulation data. This is significant, because if the electro-mechanical efficiency of the ESC-motor system is known the airspeed model can be initially calibrated using simulation data only, thereby enabling a safe initial flight. Subsequently, the in-flight identification procedure can be used to fine-tune the model coefficients for improved accuracy. For smoother real-time implementation, a \gls{rls} algorithm can instead be utilized to identify the model coefficients from \cref{eq:gs_lr}.

\begin{table*}[h]
    \centering

    \caption{Model coefficients.}
    \label{tab:coeffs}
    \begin{tabular}{l
                    S[table-format=1.3e-1] 
                    S[table-format=2.2e2]
                    S[table-format=1.3]
                    S[table-format=2.2]
                    S[table-format=2.2e3]}
    \toprule
        Training dataset & {$\beta_1$} & {$\beta_2$} & {$\alpha_0$} & {$\alpha_1$} & {$\alpha_2$} \\
    \midrule
        \gls{bem}           & 2.74e-2  & -9.91e11 & 0.869 & -3.60  & -8.18e3 \\
        Wind tunnel         & 2.63e-2  & -7.82e11 & 0.940 & -5.86  & -2.79e3 \\
        Flight              & 2.55e-2  & -7.11e11 & 0.850 & -3.87  & -4.68e3 \\
        Flight (GPS)        & 2.55e-2  & -6.85e11 & 1.180 & -13.00 & 10.30e3 \\
    \bottomrule
    \end{tabular}

    \vspace{1.5em}  

    \caption{Fitting and prediction accuracy.}
    \label{tab:pred}
    \begin{tabular}{clcccc}
        \toprule
        \multirow{2}{*}{Model} & \multirow{2}{*}{Training dataset} 
        & \multicolumn{2}{c}{\makecell{Fitting accuracy on\\ the training dataset}} 
        & \multicolumn{2}{c}{\makecell{Prediction accuracy\\ on the test dataset}} \\
        \cmidrule(lr){3-4} \cmidrule(lr){5-6}
        {} & {} 
        & nRMSE & RMSE [\si[per-mode=symbol]{\meter\per\second}] 
        & nRMSE & RMSE [\si[per-mode=symbol]{\meter\per\second}] \\
        \midrule

        \multirow{4}{*}{\parbox{3.8cm}{\centering $\beta_1 \omega + \beta_2 \dfrac{P^2}{\omega^5}$}} 
            & \gls{bem}                & 0.024 & 0.72 & 0.085 & 0.88 \\
            & Wind tunnel             & 0.095 & 0.77 & 0.057 & 0.59 \\
            & Flight                  & 0.046 & 0.84 & 0.051 & 0.53 \\
            & Flight (GPS)           & 0.038 & 0.94 & 0.051 & 0.53 \\
        \midrule
        \multirow{4}{*}{\parbox{3.2cm}{\centering $\dfrac{\omega}{2\pi}D(\alpha_0 + \alpha_1 C_P + \alpha_2 C_P^4)$}} 
            & \gls{bem}                  & 0.016 & 0.48 & 0.069 & 0.72 \\
            & Wind tunnel                 & 0.097 & 0.78 & 0.056 & 0.58 \\
            & Flight                      & 0.046 & 0.84 & 0.051 & 0.53 \\
            & Flight (GPS)           & 0.056 & 1.39 & 0.094 & 0.97 \\
        \bottomrule
    \end{tabular}

\end{table*}

\section{CONCLUSION AND FUTURE WORK}
In this work, we demonstrated that it is possible to estimate the airspeed of a \gls{uav} using only propeller power and rotational speed measurements, without relying on a vehicle model or computationally intensive algorithms. By comparing the two model structures, we found that the direct model is better suited for deployment, offering higher predictive performance when its coefficients are identified using the in-flight identification procedure. This makes the direct model well-suited for autonomous deployment.

While deriving the explicit airspeed model, we considered the effect of the time derivative of the propeller rotational speed, $\dot{\omega}$, but the optimization method did not select any terms involving it. Although an accelerating propeller implies more motor torque and thus more propeller power, its effect appears negligible in our experimental setup due to the propeller's low moment of inertia. However, this may become relevant for larger \glspl{uav} with bigger and heavier propellers.

We studied the application of the suggested method for the fixed-wing \gls{uav} category; however, the approach is not inherently bound to any specific vehicle type.
Since it relies only on the propeller operation measurements, future work may focus on relaxing the assumption of axial freestream and investigating the low advance ratio regime.
Addressing these limitations would enhance the generality of the proposed model and broaden its applicability across all \glspl{uav}.

\printbibliography

\end{document}